\documentclass[12pt]{amsart}

\usepackage{amsmath,amssymb,enumerate}

\usepackage{epsfig,fancyhdr,color}

\usepackage{amssymb}
\usepackage{amsmath,amsthm}
\usepackage{latexsym}
\usepackage{amscd}
\usepackage{psfrag}
\usepackage{graphicx}
\usepackage[latin1]{inputenc}
\usepackage[all]{xy}

\definecolor{NoteColor}{rgb}{1,0,0}

\newtheorem{ansatz}{\rm\bf Ansatz}[section]

\newtheorem{theorem}{\rm\bf Theorem}[section]

\newtheorem*{theorem 1}{\rm\bf Proposition 1}
\newtheorem*{theorem 2}{\rm\bf Proposition 2}

\theoremstyle{definition}
\newtheorem{definition}[theorem]{\rm\bf Definition}

\theoremstyle{remark}
\newtheorem{remark}[theorem]{\rm\bf Remark}

\def\interieur#1{\mathord{\mathop{\kern 0pt #1}\limits^\circ}}

\title[Geometry of Morphogenesis]{Geometry of Morphogenesis}

\author{Nadya Morozova}
\address{Insitut des Hautes \'Etudes Scientifiques\\
38 route des Chartres, 91440 Bures-sur-Yvette France\\
and Laboratoire Epig\'en\'etique et Cancer, CEA Saclay, Route Nationale, 90400 Gif-sur-Yvette France}
\email{morozova{\char'100}ihes.fr, morozova{\char'100}vjf.cnrs.fr}

\author{Robert Penner}
\address{Insitut des Hautes \'Etudes Scientifiques\\
38 route des Chartres, 91440 Bures-sur-Yvette France\\
and Math and Physics Departments, Caltech\\
Pasadena, CA 91125 USA}
\email{rpenner{\char'100}ihes.fr, rpenner{\char'100}caltech.edu}

\thanks{Thanks to Christophe Soul\'e and Sasha Getmanenko for helpful discussions.}

\begin{document}

\maketitle

\begin{abstract}
We introduce a formalism for the geometry of eukaryotic cells and organisms.
Cells are taken to be star-convex with good biological reason.  This allows for a convenient description
of their extent in space as well as all manner of cell surface gradients.  We assume that a spectrum of
such cell surface markers determines an epigenetic code for organism shape.
The union of cells in space at a moment in time is
by definition the organism  taken as a metric subspace of Euclidean space, which can be  further equipped with
an arbitrary measure.  
Each cell determines a point in space thus assigning a finite configuration of distinct points in space to an organism, and a bundle over this configuration
space is introduced with fiber a Hilbert space 
recording specific epigenetic data.  On this bundle, a Lagrangian formulation of morphogenetic dynamics is proposed based on Gromov-Hausdorff distance which at once describes both embryo development and regenerative growth.
\end{abstract}




\section{Introduction}

We seek to posit a geochemical model for morphogenesis of eukaryotic 
organisms that includes cellular details. Our aim is to describe the main laws underlying the morphogenetic processes during normal development as well as during possible perturbations resulting in creating structure de novo as in the course of regeneration \cite{lobo}.  
In this inherently multi-scale context, we must
describe the nano-biology of the cell in its environment without unnecessary detail in a manner
compatible with macro-biological aspects of the geometrical shape
of an organ or entire organism.  

Indeed, we shall proceed with a series of ans\"atze, one after another, the first group of which
are effectively axiomatic mathematical statements about the structure of cells and the organisms which they comprise.
Some aspects of this are inspired by biology as we shall discuss in detail, and others are nothing more than rather obvious but new mathematical formalizations of elementary biological structure.  Indeed, for the mathematician and biologist reader alike, the depth of
certain ans\"atze may range from sublime to ridiculous--though likely quite differently for
the two fields as we shall comment in specific cases.

The second group of ans\"atze governs the ``morphogenetic field'', a term used by Ren\'e Thom \cite{thom}
with the concept carefully articulated by the first-named author with Misha Shubin in \cite{MS} 
and discussed in some detail here, which determines the time-evolution of cell states.  
In fact, our treatment here differs from \cite{MS}  insofar as they consider four discrete cell events  (apoptosis, division, movement, growth), whereas in our language,
the first two are represented by birth/death events (death alone or death with the birth of two daughters), and the latter two by continuous evolution as a path in a certain bundle (a Hilbert space\footnote{A Hilbert space
is a vector space with inner product which is complete as a metric space, and a product
of such is thus again a Hilbert space.} as fiber over a configuration of finitely many distinct points in three-space).

One profound contribution of \cite{MS} is what we shall call the ``Epigenetic Hypothesis'':
{\it the distribution of oligo sacchyride residues of glycoconjugates on 
the cell surface determines the morphogenetic field via an as-yet unknown code.}
In fact, we are not so absolute as that and relax it here by allowing any other
distribution or gradient on the cell surface as well, notably protein or electromagnetic gradients for example.
One nice aspect of the mathematical formalism we have found is that such gradients on the cell surface
can be explicitly prescribed in terms of real-valued functions on the two-dimensional sphere.

Another notable aspect of the formalism here is the seamless jump in scale from cells to organisms provided by any
Gromov-Hausdorff  type metric on measured metric spaces.  Indeed, in our description, each cell has its explicit extent
in three-space, the union of which forms a metric subspace comprising the organism.   Depending upon which features of the organism are to be emphasized, different measures on the cells yield different metrics on the space of organisms.

Our precarious multi-scale endeavor is only confounded by the first ansatz:
\begin{ansatz} \label{ansatz1}
No statement in biology is always true except this one.
\end{ansatz}

\section{Cells}\label{sec:cells}
We first describe the geometry of a single cell in mathematical detail
(based on the biologically sound assumption that cells are star-convex),
then the notion of cell state which captures certain epigenetic information
(and is succinctly described by a collection of real-valued functions 
on the two-dimensional sphere ${\mathbb S}^2$ of unit radius centered at the origin in
space ${\mathbb R}^3$ owing to star-convexity) 
and finally discuss cell trajectory, i.e., lifetime from birth to division or to final
apoptosis (programmed cell death) or necrosis (death from injury or disease).

The structure of a  live cell includes the {\it Microtubule
Organizing Center} (MTOC) or {\it centrosome}, from which emanates a network of
microtubules forming the essential part of the cytoskeleton of the cell and holding in
place  the cell surface just as a system of poles holds in place the surface of a circus tent.
Most animal and some plant MTOC, apart from its main gamma-tubulin content, 
contain two cylindrical structures called centrioles meeting at a point; see Figure 1.


\begin{figure}[ht]
 \centerline{\epsfxsize=2in\epsfbox{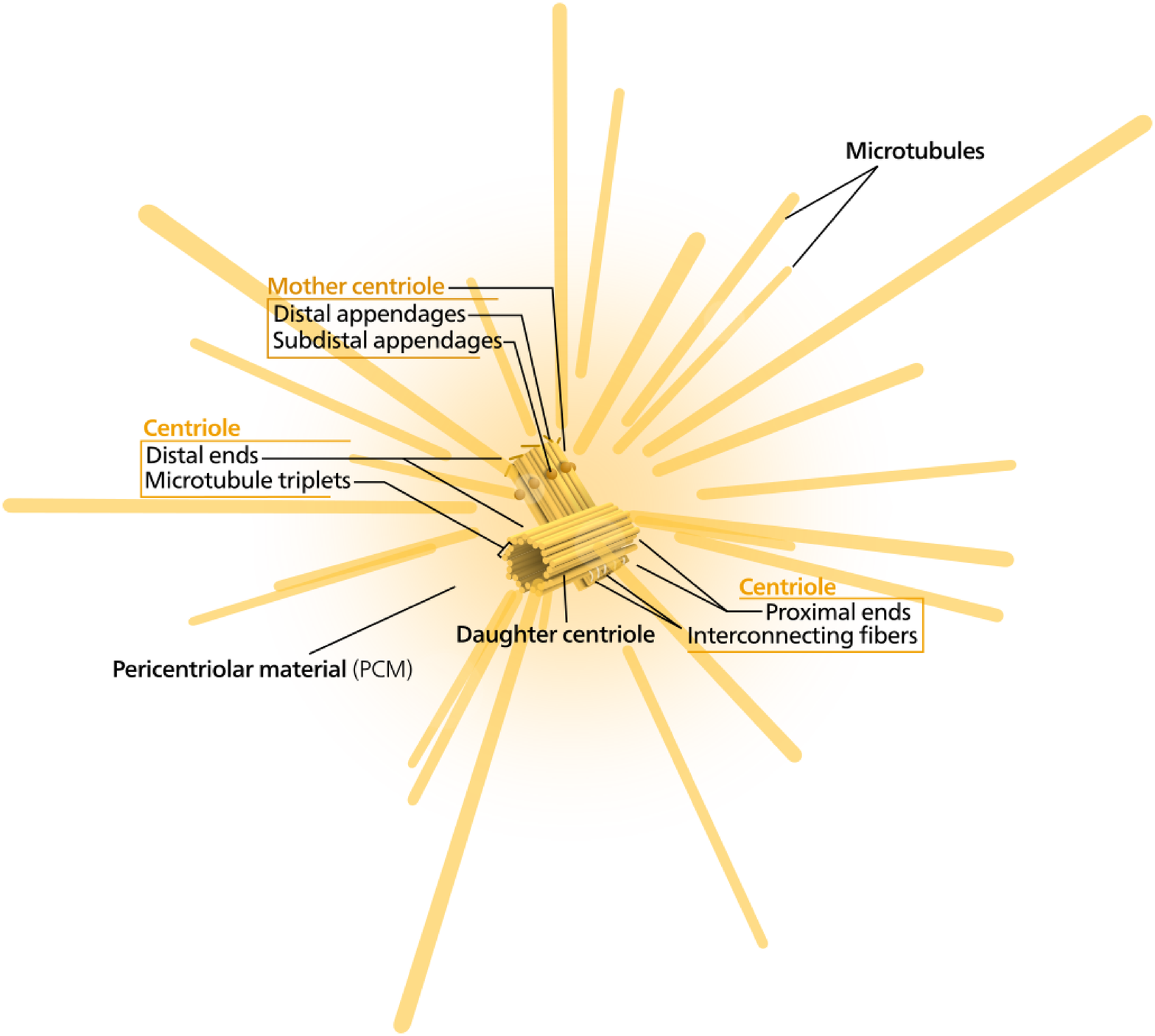}~~~~~\epsfxsize=2in\epsfbox{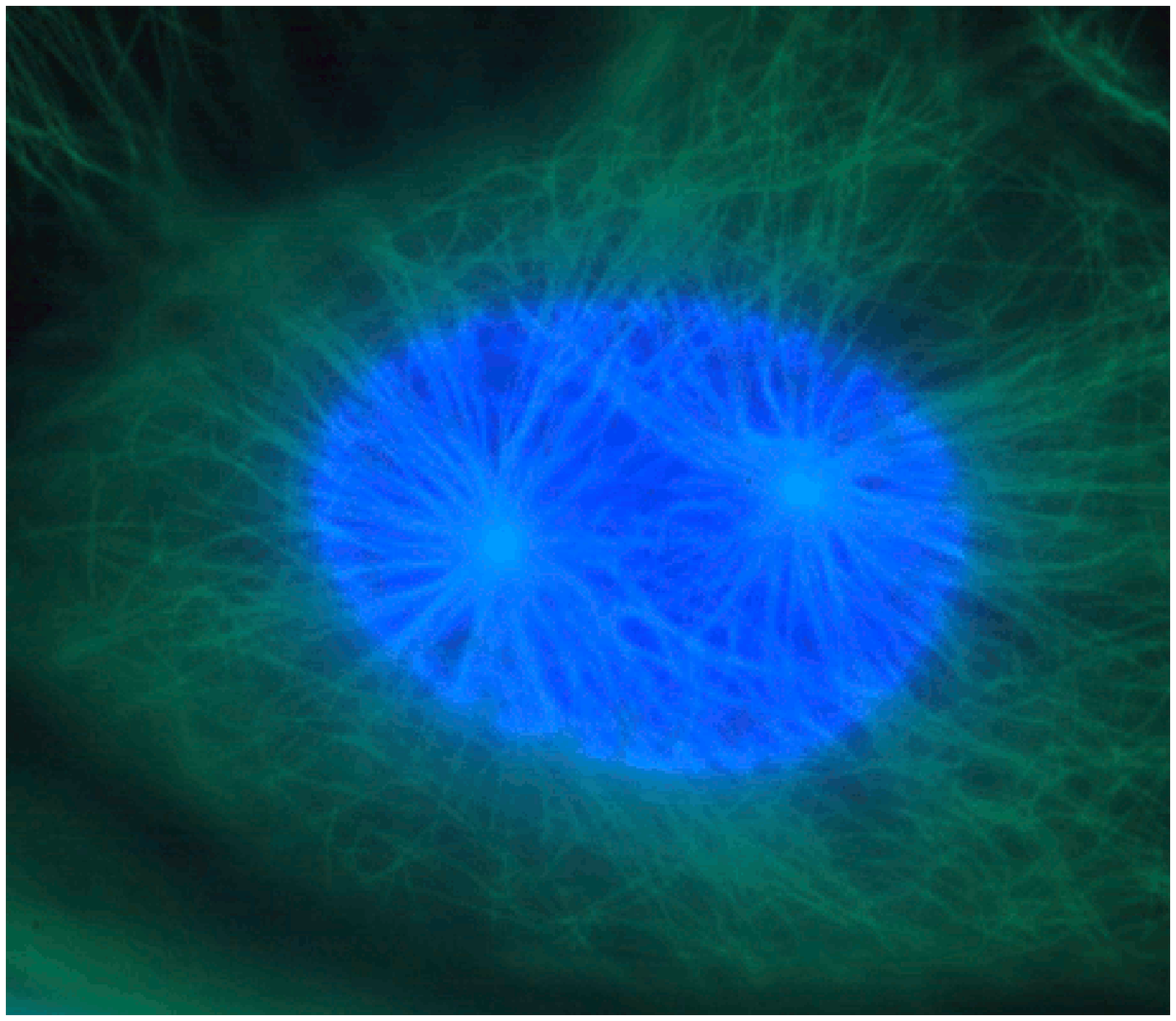}}  
  \caption{Centrosome structure (from Wikipedia) on the left and the image of centrosomes in a cell (from wadsworth.org) on the right}
 \label{fig:centro}
\end{figure}

Let us say that a cell is {\it centriolic} if its centrosome
indeed contains such a pair of centrioles, and likewise an organism if its cells are centriolic.
These organisms are especially amenable to our methods and analyses here since each cell $C$ has its distinguished {\it MTOC point} $p(C)\in C\subset {\mathbb R}^3$ given by the
point of contact of the two centrioles.  This is a mere convenience though, and one may rather make the choice of an appropriate point within the centrosome or even an arbitrary point of star-convexity in general.

\subsection{Shape} \label{sec:shapes} 

Even comprehending just the geometry
of a single cell within an organism already seems hopeless as suggested by Figure 2.
One has only to peruse for example protist organisms (which are unicellular eukaryotes)
depicted on the left to see a wild geometric diversity of detailed cell shapes.  There are furthermore
anatomical anomalies such as meter-long human neuro skeletal cells as depicted on the right with 
their dendritic extremities.  Exotic examples abound.

Nevertheless modulo Ansatz \ref{ansatz1}, all cells do share a common geometric feature as follows. 
Recall that in mathematics, one says that a region $K\subseteq {\mathbb R}^3$ of space 
${\mathbb R}^3$ is
{\it star-convex} with respect to a point $p\in K$ provided that for each
point $q\in K$ the line segment $\overline{pq}$ connecting $p$ to $q$
also lies in $K$.

\begin{figure}[ht]
 \centerline{\epsfxsize=1.9in\epsfbox{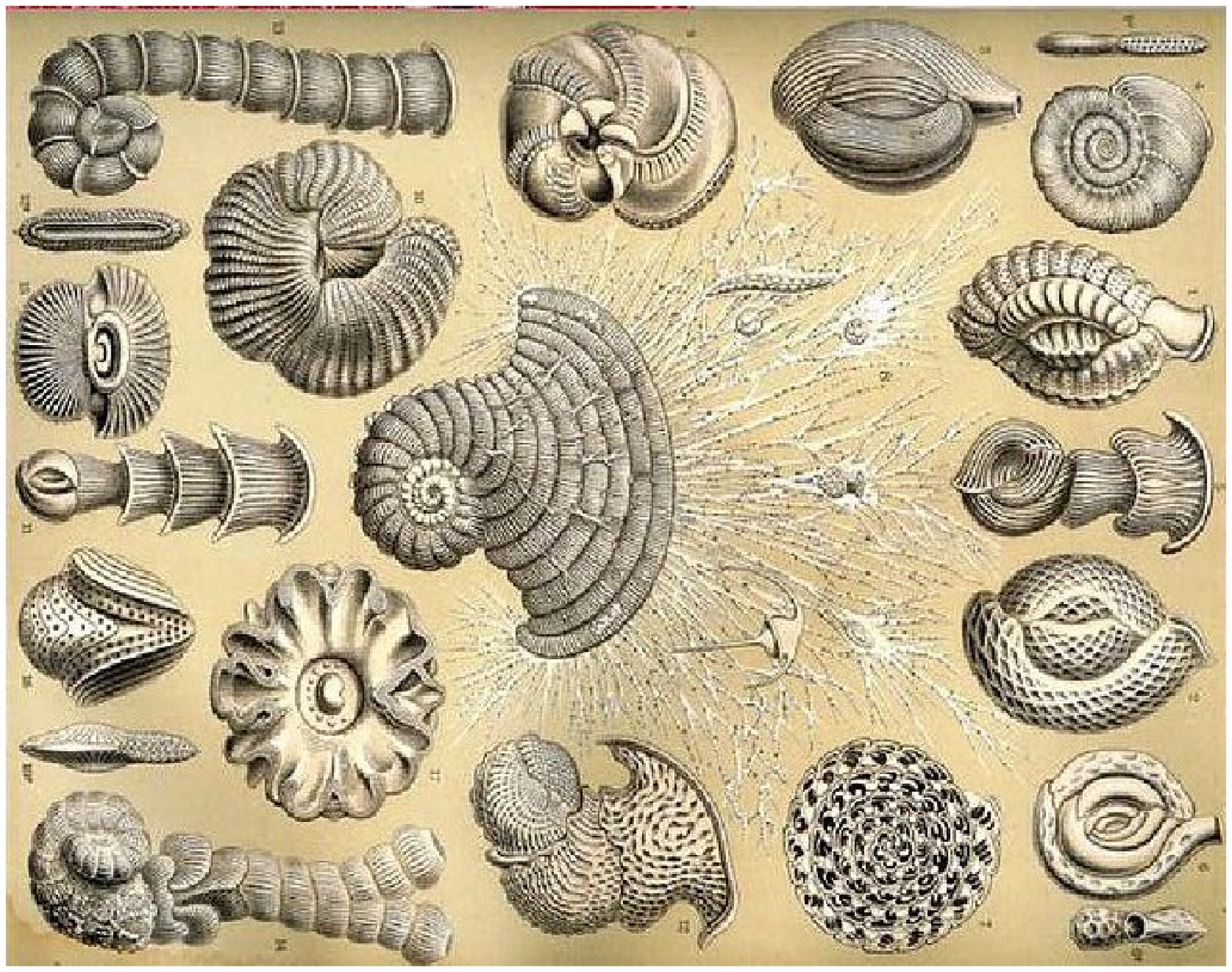}~~\epsfxsize=2.5in\epsfbox{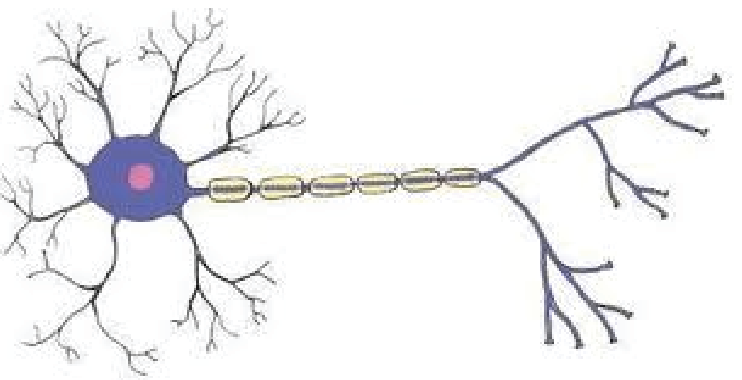}}  
 \caption{Cell shapes: collage of protists from \cite{Watson} on the left and a sketch of a multipolar neuron (from studyblue.com) on the right.}
 \label{fig:protists}%
\end{figure}

\begin{ansatz}\label{ansatz2} 
A cell $C$ is star-convex with respect to some point $p(C)\in C$ in its MTOC, and in particular
for a centriolic cell, we may take $p(C)$ as the point of contact between centrioles.
\end{ansatz}

It is a biological axiom, to the extent that such there be
in light of Ansatz \ref{ansatz1}, that the cytoskeleton of a cell is star-convex with respect to a point in its MTOC.
In our idealization of the geometry, we may consider the outer surface of the cytoskeleton 
as a cell surface.  
In reality, one would presumably take a small epsilon neighborhood of the star-convex cytoskeleton to 
describe the cell shape, though many cells are star-convex already with epsilon equal to zero.
Indeed, we shall later consider only epsilon neighborhoods of cells anyway, so the geometric
idealization of Ansatz \ref{ansatz2} from cytoskeleton to cell is unimportant.

\begin{remark}
Fix in space the centrosome of a centriolic cell at some instant in time.  Let $\vec u$ be the unit vector
in the direction along the cylindrical axis for the earlier-replicated centriole and $\vec v$ the unit vector
in the direction of the later-replicated one, with $\vec w=\vec u\times \vec v$.  In each case, the direction
along the cylindrical axis is determined by the fact that the MTOC point of centriole contact occurs asymmetrically at
a preferred starting end of each cylinder.  A centriolic cell thus determines a so-called positively oriented orthonormal
3-frame $(\vec u,\vec v,\vec w)$ describing the spatial orientation of each cell.  We shall not in fact require this further extra structure,
but it is worth noting.
\end{remark}

Now, fix a cell $C$ with its MTOC point $p=p_C$.  Given a unit vector $\vec \xi\in{\mathbb S}^2$, where ${\mathbb S}^2$ denotes the usual two-dimensional
sphere in ${\mathbb R}^3$ (comprised of endpoints of vectors of unit length based at the origin), there is a corresponding ray
from $p$ in the same direction of $\vec \xi$, and this ray meets the cell surface at a single point at some distance
$\rho(\vec \xi)\in {\mathbb R}_{> 0}$ from $p$ according to star-convexity.  See Figure 3.
This assignment $\rho=\rho_C:{\mathbb S}^2\to{\mathbb R}_{> 0}$ of a positive real-valued function on the two-dimensional sphere
${\mathbb S}^2$ completely determines the shape of the cell $C$.

\begin{figure}[ht]
 \centerline{\epsfxsize=2in\epsfbox{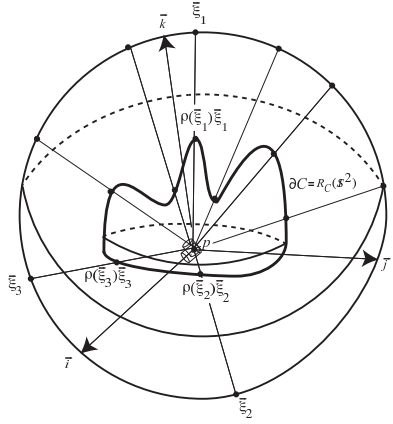}}  
 \caption{Centrosome and shape function $\rho$.}
 \label{fig:frames}%
\end{figure}

To recapitulate this section, given a eukaryotic cell $C$, there is an MTOC point $p=p_C\in C$,
which is well-defined without choice provided $C$ is centriolic, with respect to which $C$ is
star-convex together with a shape function $\rho=\rho_C:{\mathbb S}^2\to {\mathbb R}_{>0}$ which completely determines the geometrical cell
shape $$C=\{ q\in {\mathbb R}^3: ||q-p||\leq \rho({{q-p}\over{||q-p||}})\}\subseteq{\mathbb R}^3,$$
where $||\vec \eta||$ denotes the usual length of a vector $\vec\eta$ in ${\mathbb R}^3$. 
There is furthermore a canonical identification of the standard sphere ${\mathbb S}^2$ with the cell surface $\partial{C}$
given by the mapping $R_C:\vec\xi\mapsto p_C+\rho_C(\vec\xi)\vec\xi$.
The volume $V$ of $C$ and its surface area $A$ are respectively given in terms of a sufficiently regular shape function $\rho_C$ as
$$\aligned
V(C)&=\int\int\int_C dV={1\over 3} \int\int_{{\mathbb S}^2} [\rho_C(\vec\xi)]^3~ dA,\cr
A(C)&=\int\int_{{\mathbb S}^2} \bigl | \nabla \rho_C(\vec\xi)\cdot\vec \xi \bigr |~dA,\cr
\endaligned$$

\subsection{Cell state}\label{sec:states}

What exactly constitutes a cell state for morphogenesis or any other purpose depends upon the implied internal structure of the cell and 
must reflect the predictions and interactions of the theory that depend upon it.  
One obvious aspect of cell state is the
shape function $\rho_C:{\mathbb S}^2\to{\mathbb R}_{>0}$ and another is
the spatial location of the
MTOC point $p_C$ introduced in the previous section.  Taken together, these
determine the natural parametrization
$R_C:{\mathbb S}^2\to\partial{C}$ of the cell surface $\partial{C}$.

It follows from the existence of the parametrization that any real-valued function $\phi:\partial{C}\to{\mathbb R}$ can equivalently be specified 
by simply pulling back
to a real-valued function $f:{\mathbb S}^2\to {\mathbb R}$, i.e., $\phi(q)=f(R_C^{-1}(q))$.  In particular, the density of some
substance on the cell surface can be conveniently specified by just such a real-valued function on ${\mathbb S}^2$
\footnote{Of course, a surface density is not simply pulled-back, rather, it is scaled by the Jacobian of $R_C$ to account for area distortion}.

\begin{ansatz} 
Cell state is determined by the shape function $f_0=\rho$ together with a collection $f_1,\ldots ,f_N$ of cell surface densities $f_i:{\mathbb S}^2\to{\mathbb R}$, for $i=1,\ldots ,N$.
\end{ansatz}

The collection $(f_1,\ldots ,f_N)$ of cell surface densities is called the {\it epigenetic spectrum}
of the the cell.  We shall refer to the full $N+1$ tuple  $F_C=(f_0,f_1,\ldots ,f_N)$  including the shape as the {\it (internal) cell state}.

In fact, the cell surface $\partial{C}$ is relatively fluid with sub-regions evolving in time to different shapes within the cell surface.  Thus, we should expect the typical distribution
of substance on the cell surface to evolve continuously in time.  

Of special interest are the distributions of oligo conjugates, small oligosaccharide residues of glycoconjugates\footnote{
The term ``glycoconjugates" means glyco-residues attached to either lipids or proteins lying inside the cell membrane or cell wall in the case of plants. }
on the cell surface. There are up to 12 varieties of monosaccharides and disaccharides in oligosaccharide residues of 
glycoconjugates (six of hexose type).  There have been numerous indications that these oligosaccharide residues are related to cellular 
morphogenetic pathways as has been formulated as an Epigenetic Hypothesis in \cite{MS} and the references therein.

Thus, we propose that the cell state $F_C$ of $C$ includes both the cell shape 
$f_0=\rho$ and the epigenetic spectrum of cell densities $f_1,f_2,\ldots , f_{12}$ of the corresponding number of ``coding" oligo conjugates, as per the Epigenetic Hypothesis, the nature and number of which should be determined experimentally.  Here is an important point of contact between one aspect of our theoretical work and experimental determination.
We leave open the possibility of extending this minimal set of functions determining cell state with other molecular compounds or other factors such as protein or electromagnetic gradients defined on the cell surface, where recent works \cite{Levin} suggest the latter are central to morphogenesis. 

Recall \cite{izzy} that there is a standard distance
 between two tuples
$F=(f_0,f_1,\ldots ,f_N)$ and $G=(g_0,g_1,\ldots ,g_N)$ of functions
$f_i,g_i:{\mathbb S}^2\to{\mathbb R}$ given by their {\it $L^2$ distance}
$$d(F,G)=\sqrt{\int_{{\mathbb S}^2} (f_0-g_0)^2+(f_1-g_1)^2+\cdots +(f_N-g_N)^2 dA}.$$
Indeed, letting 
$$L^2({\mathbb S}^2)=\{ f:{\mathbb S}^2\to{\mathbb R}:\int\int_{{\mathbb S}^2} f^2 dA <\infty\}$$
denote the so-called square integrable functions on ${\mathbb S}^2$, the distance $d$ gives the collection
$\bigl [ L^2({\mathbb S}^2)\bigr ]^{N+1}$ of all possible cell states $F_C$ the structure of a Hilbert space.  Hardly worth mentioning mathematically, the next ansatz gives an explicit concept of distance between cell states.

\begin{ansatz}\label{a:L^2}
The natural distance\footnote{In fact, we mean this in the looser sense that the distance might differently weight different
component functions (a ``weighted $L^2$ space'') more generally taking $d_{\vec\alpha}(F,G)=\sqrt{\int\int_{{\mathbb S}^2} \sum_{i=0}^N \alpha_i (f_i-g_i)^2 dA}$, for
certain specified parameters $\alpha_i\in{\mathbb R}_{>0}$. }
 between cell states $F=(f_0,f_1,\ldots ,f_N)$ and $G=(g_0,g_1,\ldots ,g_N)$
is the $L^2$ distance $d(F,G)$.
\end{ansatz}

\subsection{Trajectory}\label{sec:life}
 The sperm-fertilized cell oocyte becomes the zygote,
the first cell $C({\emptyset})$ of the organism, which next 
 mitotically divides to ten or more generations in a highly organized paroxysm of cell divisions, 
no time at all for diffusive or other equilibria to develop, no time for appreciable cell growth or motion and immediately generating all manner of steep gradients
across the early embryo.

Recall that the centrosome of a cell $C$ at interphase
replicates to create a pair of centrosomes.  These  centrosomes move to opposite ends of the cell and nucleate microtubules to help form the mitotic spindle separating the chromosomes into two sets, one copy for each daughter cell $C',C''$ together with one centrosome for each daughter cell.
One centrosome of $C'$ or $C''$ is inherited from the centrosome of $C$, which we call the mother centrosome, while the other, which we call the daughter centrosome since it is younger than its mother, has been assembled upon the mother centrosome before mitosis. 

Insofar as the replicated centrosomes of $C$ are distinguishable, so too are its daughters $C'$ and $C''$. We can thus
label each cell in the organism by its phylogeny: a word in the alphabet $m$ and $d$, where $m$ indicates the daughter  $C(m)$ that
inherits the mother centrosome from $C$ and $d$ the daughter $C(d)$ that inherits the daughter centrosome.

Each cell itself is determined by its {\it phylogeny}, a finite word $\omega$ in the two-letter alphabet $\{ m,d\}$, 
including the empty word for the zygote.  This labeling of cells by $m$ and $d$ is evidently equivalent to the usual structure of  
embedded phylogenetic tree: a binary tree with gamete for root and with $m$ on the right (say) and $d$ on the left.  

\begin{ansatz}\label{trajectory}
Each cell $C$ is born at some instant $s_C$ in time and terminates either through death or cell division at a
later instant $t_C$; each cell except for the gamete is born with exactly one distinguishable sister.  
\end{ansatz}

Death may be apoptosis, necrosis or it may reflect some crisis for the organism such as  amputation or disintegration.  A somatic cell is always born with exactly one sister 
and terminates either alone (by death) or in giving birth to its daughters (by division).

It is worth saying explicitly that Ansatz \ref{trajectory} is biologically tautological, no more than a self-evident remark but is a necessary aspect of axiomatic mathematical formulation.  Indeed, in other important classical mathematical contexts, births as well as deaths can occur only in pairs.  Biologically, the ansatz is trivial and can be ignored, for of course it is tautologically so.

Vertices of our phylogenetic tree other than the root are thus labeled by the cell event of death or cell division.  Edges of the tree correspond to the lifetime of cells, and we will assign as a length of the edge labeled by the cell $C$ its lifetime $t_C-s_C$.  In this manner, a point in an edge corresponds to the cell $C^t\subseteq {\mathbb R}^3$ at a particular instant $t$ in time, where  $s_C\leq t\leq t_C$.

According to our assumptions, two sister cells $C'$ and $C''$ must share their
time of birth $s_{C'}=s_{C''}=t_C$ with the time of death of their mother (but see section \ref{sec:gap}).  
For an extreme example such as a skeletal muscle cell $C$ or an eye lens cell $C$ which is fully differentiated
once and for all time already in the embryo and never divides again, we find $s_C$ at birth and $t_C$ after death of the organism.  At another extreme are dermal cells with their profuse and nearly identical divisions.

The potency of a cell to produce progeny with different types of tissues and organs or simply to differentiate into many possible cell types is called a morphogenetic potency or just a potency.  Thus in an organism there are totipotent, pluripotent, bipotent and fully differentiated cells. 

Let us define the {\it potency} $\pi(C)$ of a cell $C$ to be zero if it does not divide at all, and if it does divide, say the daughters
have respective epigenetic spectra $F',F''\in[L^2({\mathbb S}^2)]^{N}$, then we compare the spectra defining
$\pi(C)=d(F',F'')$.  Thus, the potency of a fully differentiated cell is nearly  zero.  In contrast, a bipotent stem cell,
for example, can have a large potency if its product cell line differs substantially in epigenetic spectrum from its stem cell line in the $L^2$ distance.

\section{Organisms}
We define an organism to be the collection of its constituent cells lying in space
as determined by a language over the two-letter alphabet given by following the centrosome.  The collection of 
constituent cell states determines a state of the organism at each
instant in time.  A convenient mathematical formalism exists for this as a
configuration space of distinct points in space which evolve under time either continuously or through
specific rules of birth and death at certain event times together with a bundle over this base
space with fiber given by an $L^2$ space of functions on the two-dimensional sphere.  All this is quite mathematically routine
given what has come before.  

At a given moment in time, the cells $C$ underlying an organism comprise a metric
subspace which moreover can be further imbued with any  function $\mu(C)$ chosen to reflect essential morphological or other parameters of $C$.
For example, we could take mass density of a cell, density of distribution of ion channels or other cell constituent concentrations.  We may choose particular functions $\mu(C)$ for specific applications

A fundamental ansatz is that two organisms each at a given moment in time can be profitably compared using Gromov-Hausdorff (GH) type distances (cf.\ Appendix A) between corresponding measured metric spaces.   This is precisely what allows us to finally pass from the cellular to the capacious scale in our investigations.

It is important to note that the comparison of ``two organisms" in this framework should be interpreted first of all as the comparison of the morphology of the same organism at different moments in time reflecting the laws of evolution of its shape.  Equally, we consider the comparison of the differences between normal and ``crisis" developmental scenarios (that is, strong
perturbations of proper morphology) 
for the same organism at some moment in time.

\subsection{Phylogeny}\label{sec:ophylo}

The zygote $C(\emptyset)$ comes into being at an instant $s_\emptyset=s_{C(\emptyset)}$ and then undergoes a sequence of
binary divisions, the first one occurring at the instant $t_\emptyset=t_{C(\emptyset)}$.  At each division, there are distinguishable sister
cells produced, and the mother cell producing these progeny subsequently ceases to exist.  And so it goes.

The mother of all cells of the organism is the zygote $C(\emptyset)$.  For each subsequent offspring produced, i.e., each cell
that did, does or will or ever occur in the organism there is a well-defined 
word $\omega$ in the two-letter alphabet $\{ m,d\}$ as discussed before, where the length of the word is the number of generations
from the zygote.  

Not all words occur for a given organism, and the {\it phylogenetic language} of ${\mathcal O}$ is the collection
$\Omega_{\mathcal O}$ of words that do in fact
occur for the organism ${\mathcal O}$.  Let $C(\omega)$ be the cell corresponding to the word $\omega\in{\Omega}_{\mathcal O}$
including the empty word for the zygote.  Thus, each cell $C(\omega)$ is born and dies at respective specific times $s_\omega$ and $t_\omega>s_\omega$.

We shall let $C^t_\omega\subseteq{\mathbb R}^3$ denote the cell $C(\omega)$ labelled by $\omega$ at time $t$,
where $s_\omega\leq t\leq t_\omega$ with time-dependent cell state $F^t_\omega\in [L^2({\mathbb S}^2)]^{N+1}$, reflecting shape and epigenetic coding, and
MTOC point $p_\omega ^t\in{\mathbb R}^3$
with auxiliary measure $\mu^t_\omega\in{\mathbb R}_{>0}$.

\subsection{Organism state, configuration and trajectory}\label{sec:otraj}

We have the mathematical definitions of organism and state given by
the following biological tautology:

\begin{ansatz}
An organism $\mathcal O$ is determined by the collection of its constituent cells $\{ C(\omega):\omega\in\Omega_{\mathcal O}\}$ and its state by their respective states.
\end{ansatz}

Let us say that a cell $C^t_\omega$ {\it exists} at some time $t$ if $s_\omega\leq t\leq t_\omega$ for some $\omega\in\Omega_{\mathcal O}$.
It is convenient to suppress the birth/death times of a cell and write simply $C^t_\omega\subset {\mathbb R}^3$ to signify the cell at a time $t$ at which the cell exists and to signify the empty set otherwise.  In the same way, let $F^t_\omega$ denote the cell state and $\mu^t_\omega$ the measure at a time $t$ between cell birth and death and to take constant value zero otherwise; finally,
$p^t_\omega$ denotes the MTOC point and $\rho_\omega^t$ the shape function if the cell
$C^t_\omega$ exists, and these are undefined otherwise.

Let $\Omega^t=\Omega^t_{\mathcal O} \subseteq \Omega_{\mathcal O}$ be the phylogenetic words of the 
cells of  the organism ${\mathcal O}$ that exist at time $t$.
The organism 
$${\mathcal O}^t=\bigcup \{C^t_\omega:\omega\in\Omega^t\}\subseteq {\mathbb R}^3$$ 
at a time $t$ is the union of these cells, i.e., the constituent cells of the organism at the instant $t$.  
Of course, ${\mathcal O}^t$ or any epsilon neighborhood of it inherits the structure of a metric subspace of ${\mathbb R}^3$,
and each cell comes with its measure $\mu^t_\omega$ which together give
the weighted sum
$$\mu^t=\mu^t_{\mathcal O}=\sum_\omega \mu^t_\omega \delta _{p^t_\omega}$$
of delta functions as a measure on ${\mathcal O}^t$.

The {\it state} of the weighted organism $({\mathcal O}^t,\mu^t)$ at time $t$ consists of the cell states $F^t_\omega$
of its constituent cells.  The collection of MTOC points 
 $$\chi^t=\{p^t_{\omega_1},p^t_{\omega _2},\ldots , p^t_{\omega_{\#\Omega ^t}}\}$$
 of cells that exist at time $t$ come in their prescribed lexicographic ordering (where, say $m<d$)
which we shall call the {\it (labelled) configuration} of the organism at time $t$.

Recall that if $X$ is a metric space, then a {\it (labelled) configuration} in $X$ of $n\geq 1$ points is a collection of $n$ distinct labelled points in $X$; see Appendix B for the barest thumbnail discussion tailored to our own needs and \cite{birman} for instance for further generalities.  It is biologically clear that in fact the MTOC points $\chi^t$ do give a configuration  in ${\mathbb R}^3$,  lying in ${\mathcal O}^t$, i.e., the MTOC points of distinct cells never coincide.  Indeed, we have the stronger:

\begin{ansatz} 
The cells of an organism that exist at each instant in time have disjoint interiors in space ${\mathbb R}^3$.
\end{ansatz}

Again a biological tautology, this kind of ``steric'' constraint is commonplace in protein theory for instance.  We do not intend to entirely rule out
gap or other cell junctions here, rather, let us consider adding that further layer of detail only later.

The number $\# \chi^t$ of MTOC points of the organism ${\mathcal O}^t$ is piecewise constant and 
jumps at various times of lone death (one point disappears) 
or birth (two new points appear and an old one disappears) of constituent cells according to Ansatz \ref{trajectory}.  In fact, there may be finitely many coincidentally simultaneous birth or death times, but we make the following finiteness assumption:

\begin{ansatz} \label{a:wha} 
For any fixed $t$, there is a {\rm finite} sequence of {\rm event times} $\tau_0=s_\emptyset<\tau_1<\tau_2<\cdots <\tau_M\leq t$
so that for each time $\tau_{i-1}<s<\tau_i$, for $i\geq 1$, the configurations $\chi^s$ are equinumerous and evolve continuously in time.
\end{ansatz}

The discussion of how exactly the cell state before determines the specific cell states after a birth event time is postponed until the more speculative considerations of the next section.

\section{Morphogenesis}\label{sec:morpho}

``The time has come'' the Walrus said,
``To talk of many things:
Of shoes--and ships--and sealing-wax--
Of cabbages--and kings--
And why the sea is boiling hot--
And whether pigs have wings.''\footnote{Lewis Carroll, The Walrus and The Carpenter (1872).}
This is just to say that now we must discuss things a bit more metaphorically in order to absorb the next ans\"atze. 

\subsection{Organism shape}\label{sec:oshape}

As humans, we can perceive the natural
world on various levels.  For instance, in pond water
one finds paramecia, which we might probe in the lab {\it in vivo}
or {\it in vitro} to discern further structure.  Going to an extreme,
we might travel to CERN and uncover the constituent elementary particles.

As human observers, we do not even want to try to absorb the molecular let alone the atomic or stringy nature of a paramecium.  With what  has been\footnote{Ren\'e Thom, Semiophysics: a sketch (1989)} called ``willful ignorance''
and also\footnote{Claude Levi-Strauss, Tristes Tropiques (1955)} termed ``twice barbarism'',  our vision in morphogenesis had better contain the required cellular and other data even if we shall ultimately and barbarically ignore this level of detail in the large.

To handle the need for minute detail in certain biological regards for cells,
we have already introduced the notion of cell state.  We must somehow pass from this
data as we have formalized it, corpuscular by its very nature, to the context of continuous data of substantial spatial extent.

Naturally enough, we might take an epsilon neighborhood of the actual cells at each instant in time, that is, a neighborhood of
${\mathcal O}^t$ in ${\mathbb R}^3$.  We furthermore might naturally let epsilon vary over ${\mathcal O}^t$ in order to allow for instance for different packing densities of one type of cell compared to another.  
An elegant and precise mathematical formalism for this viewpoint is provided by the class of metrics on measured metric spaces inspired by GH distance.  For completeness, one such distance (the Sturm $L^2$ distance \cite{Stu06}) is explicitly defined in Appendix A.

\begin{ansatz}\label{ans:gh} 
Given two organisms ${\mathcal O}$ and ${\mathcal O}'$ with their respective measures $\mu$ and $\mu'$, Gromov-Hausdorff type distances between the measured metric spaces $({\mathcal O},\mu)$ and $({\mathcal O}',\mu')$
such as the Sturm $L^2$ distance capture the geometry of organism morphology.
\end{ansatz}

Just as the Sturm $L^2$ distance is one among many sensible choices for comparing measured metric spaces, so too, we have the flexibility to take different measures $\mu(C)$ on cells $C$.  As an example, the mass density can be one special case of measures involved in building the morphology of an organism.  Presumably as a human observer employing it as we have done here is like observing the world with those X-ray glasses that are bought by gullible children; we would detect density differences between cells within organisms when we compare them wearing our mass density glasses.  A tadpole and a frog would be easily distinguished by their bone structures for instance.  The formalism we have developed is robust enough to handle general examples exchanging X-ray for other glasses by changing the measure $\mu^t$ on the fixed metric space that is the organism ${\mathcal O}^t$.

\subsection{Cell division in detail}\label{sec:gap}

During mitotic division of the cell $C$, the two daughter cells $C(m)$ and $C(d)$ labelled by centrosomes as before are separated by a so-called mitotic plane passing through the interior of $C$ often asymmetrically. This separating plane divides the cell surface $\partial{C}$ into two inequivalent regions $\Omega_m,\Omega_d\subset\partial{C}$
forming the cell boundary of respective daughters.

\begin{ansatz}\label{a:mbe} 
The epigenetic spectra of daughters $C(m)$ and $C(d)$
are nearly
inherited in the $L^2$ sense on ${\mathbb S}^2$ by inclusion of the cell surfaces $\Omega_m,\Omega_d$ from the mother.
\end{ansatz}

On the complementary regions to $\Omega_m$ and $\Omega _d$ in daughter cell surfaces, there are as-yet unspecified rules of how to extend to a function on the entire cell surface.
This ansatz partly describes the {\it morphogenetic birth effect}, that is, 
the function that determines the cell states of $C(m)^{\sigma_2}$ and $C(d)^{\sigma _2}$ at the moment $\sigma_2$ when mitotis is complete 
from the cell state of $C^{\sigma_1}$  of a mother cell as mitosis begins at $\sigma_1$. 

It is not that the biology breaks down during the interval between times $\sigma_1$ and $\sigma_2$.  After all, there is a sensible moment of 
birth when the cell membranes of the daughters finally separate that we could also take to be the moment of death of the mother $C$.
The surface concentrations in $\partial C$ of the oligo conjugates or other compounds are everywhere defined until this moment
and after become the surface concentrations in $\partial C(m)$ and $\partial C(d)$.
 In this sense, again the ansatz is disappointingly tautological biologically.  However,
this scenario does {\sl not hold exactly} and in our barbaric and willful ignorance, we posit in Ansatz \ref{a:mbe} a precise sense in which $L^2$ distance suitably measures deviation from this idealized biology.

\subsection{Morphogenetic field}

To recapitulate, given a eukaryotic  organism ${\mathcal O}$ and a time $t$, we have the constituent cells 
$C_\omega^t\subseteq {\mathbb R}^3$ that exist at time $t$ indexed by a finite set $\Omega^t$ of words in the alphabet 
$\{ m,d\}$.  The union of these star-convex regions is by definition the organism ${\mathcal O}^t\subseteq{\mathbb R}^3$
at time $t$ as a metric subspace of ${\mathbb R}^3$.  This region comes equipped with a measure $\mu^t$,
and it is the measured metric spaces $({\mathcal O}^t,\mu^t)$ that reflect the large-scale geometry of the organism.
Standard methods, called GH type metrics, can be applied to these spaces to quantify the sense in which two organisms have the same shape.  Nevertheless, the organism has a complicated state,
namely, we take the state of the organism to be the collection of cell states of each of its constituent cells,
i.e., cell states $F_\omega^t$ and MTOC points $\chi^t_\omega$, for $\omega\in\Omega^t$

On the cellular scale, we have organized things in Appendix B into a vector bundle $E\to \Gamma$ over the
space $\Gamma$ of configurations of finitely many points in ${\mathbb R}^3$, where the fiber over a configuration
with $n$ points is an $n$-tuple of elements in $[L^2({\mathbb S}^2)]^{N+1}$, namely, the cell states
of the constituent cells.  Thus, the entire organism in its entire state is described at an instant in time
as a point in the total space of this bundle over the finite configuration of MTOC points
in the base, and the measure $\mu$ is an auxiliary function defined on the base.

The idea from \cite{MS} is that the ``morphogenetic field'' is a function defined on some bundle of states
that governs the time evolution of an organism starting from ovum.  
Their formalism was different with discrete time cell events as was already mentioned.  Here we
imagine continuous evolution in some $E^{(n)}$ with jumps in $n$ at finitely many event times compatible with Ansatz
\ref{a:wha}.  The morphogenetic field of \cite{MS} should be realized here as some sort of functional which governs the time evolution of an organism, namely, as a Lagrangian action:  

\begin{ansatz}\label{a:lag} 
There is some Lagrangian formulation of organism evolution, i.e., an action
functional on paths in $E$ whose stationary paths
are the time trajectories of organisms.
\end{ansatz}

It is important to emphasize that the Lagrangian formulation of Ansatz 4.3 is
different from that of the underlying physics: our action should be formulated on the space of organism states as we have defined them.

Using our definition of potency (see the end of section \ref{sec:life}) and the observation that each organism starts to differentiate from the zygote having a huge potency towards its later less-potent progeny, we posit: 

\begin{ansatz} 
The average {\it potency} $\pi(C)$ over the set of all cells $C$ in an organism is strictly decreasing in generation during embryo development.
\end{ansatz}

This is of course in keeping with the second law of thermodynamics.

\begin{ansatz} \label{a:sis} 
Nearly identical sisters almost always remain nearly in contact.
\end{ansatz}

This is interesting as it relates the phylogeny to the geometry.  Of course, sisters are never
truly identical, but if they were, then the morphogenetic field could not distinguish them,
so of course they would evolve in the same manner.  Nature is nearly like this as an essentially deterministic physical system.  This seems to give a nice explanation for organ formation and other
aspects as well and really makes good sense.  Again, with the $L^2$ distance between cell states, we can conveniently quantify differences between sisters or indeed any pair of cells viz Ansatz \ref{a:sis}

\begin{ansatz} 
The morphogenetic field is locally supported in the sense that for any cell $C^t_\omega$ that exists, there is a neighborhood
${\mathcal N}\supseteq C^t_\omega$ of it in ${\mathbb R}^3$ so that the forward time evolution of $C^t_\omega$ for some fixed interval of time depends only upon the states of the cells in ${\mathcal N}$.
\end{ansatz}

In particular, phylogeny does not matter {\it a priori}, which implies
that there is indeed a physical determinism underlying the morphological determinism we are studying.  And this determinism is governed by the laws formulated as a morphogenetic field concept.

\begin{ansatz}\label{a:stable}
For any sub-organism ${\mathcal O}'\subseteq{\mathcal O}^t$ in any reasonable state there is a stable ideal outcome organism
${\mathcal O}^\infty$ so that the Lagrangian action of an arbtrary path from
${\mathcal O}'$ to a stable outcome ${\mathcal O}^*$ is given by a Gromov-Hausdorff
type distance between ${\mathcal O}^\infty$ and ${\mathcal O}^*$.
\end{ansatz}

This GH-deviation from ideal outcome is admittedly a bold hypothesis for the dynamics
providing thereby the key mechanism driving morphogenesis ranging from embryo development to the creation of  proper morphology during the regeneration processes in the event of crisis.  
The ansatz is interesting on several levels first of all in its postulate that there are stable ideal outcome
organisms at all under admittedly unspecified reasonable conditions.  The comparison 
of stable outcomes using GH techniques is consistent with the
earlier Ansatz \ref{ans:gh}.
Indeed, perhaps another more traditional differential calculus type expression of action
in terms of the GH-metric could be possible rather than relying on stability of outcomes.
Secondly, it ties the cellular scale of the morphogenetic field with
the large-scale geometry of organisms through the action, and it evidently
implies the earlier Ansatz \ref{a:lag} in its specificity.

\appendix\relax

\section{Distances between measured metric spaces}\label{app:a}

This appendix gives a precise example of the sort of distance between
measured metric spaces that we propose for use in comparing organisms.  For completeness, definitions
are given from first principles.

\begin{definition} \label{def:metric}If $X$ is a set, then a function $\mu:X\times X\to{\mathbb R}$ is a {\it metric} if and only if
for all $x_1,x_2,x_3\in X$, we have:
{ [Symmetric]} $d(x_1,x_2)=d(x_2,x_1)$;
 {[Triangle Inequality]} $d(x_1,x_3)\leq d(x_1,x_2)+d(x_2,x_3)$;
{[Positive]} $d(x_1,x_2)\geq 0$ and $d(x_1,x_2)=0$ if and only if $x_1=x_2$.

\end{definition}

Given metric spaces $(X,d_X)$ and $(Y,d_Y)$,
let ${\mathcal D}(d_X,d_Y)$ denote the set of all possible metrics on the
disjoint union $X\sqcup Y$ of $X$ and $Y$ extending $d_X$ on $X$ and 
$d_Y$ on $Y$, that is, $d$ satisfies Definition \ref{def:metric} on $X\sqcup Y$ and
$$d(x_1,x_2)=d_X(x_1,x_2)~{\rm and}~d(y_1,y_2)=d_Y(y_1,y_2)$$ for all
$x_i\in X$ and $y_i\in Y$, for $i=1,2$.

\begin{definition}\label{d:bpm} A {\it Borel probability measure} $\mu$ on a space $X$ is the specification of a real number $\mu(A)$ to each set $A$ in the sigma algebra\footnote{A collection of subsets of $X$ is a {\it sigma algebra} provided
it is closed under countably many applications of the set operations complement, union and intersection.
}
 ${\mathcal B}(X)$ of subsets generated by the open sets of $X$ satisfying the following properties:
{ [Sigma additive]} if $A_i\in{\mathcal B}(X)$ are pairwise disjoint
subsets for $i\in I$ with $I$ countable, then $\mu(\cup\{ A_i:i\in I\})=\sum _{i\in I}\mu(A_i)$;
{[Non-negative]} $\mu(A)\geq 0$~for all $A\in{\mathcal B}(X)$;
{[Null empty set]} $\mu(\emptyset)=0$.
{[Probability measure]} $\mu(X)=1$.
\end{definition}

Given two measured metric spaces $(X,d_X,\mu_X)$
and $(Y,d_Y,\mu_Y)$,  let ${\mathcal M}(\mu_X,\mu_Y)$ denote the
collection of all Borel probability measures
$\mu$ on $X\times Y$ extending $\mu_X$ on $X$ and $\mu_Y$ on $Y$, that is, $\mu$ satisfies
Definition \ref{d:bpm} on ${\mathcal B}(X\times Y)$ and
$$\mu(A\times Y)=\mu_X(A)~{\rm and}~\mu(X\times B)=\mu_Y(B)$$
for all $A\in{\mathcal B}(X)$, $B\in{\mathcal B}(Y)$.

One collection of distances  for measured metric spaces inspired by the GH
metric is given by:

\begin{definition}
The {\it Sturm $L^p$-distance} \cite{Stu06} between measured metric spaces $(X,d_X,\mu_X)$ 
and $(Y,d_Y,\mu_Y)$ is defined to be
$$S_p(X,Y)=\inf_{d,\mu}\biggl ( \int_{X\times Y} [d(x,y)]^p \mu(dx,dy)\biggr )^{1\over p},$$
where the infimum is over all $d\in{\mathcal D}(d_X,d_Y)$ and $\mu\in{\mathcal M}(\mu_X,\mu_Y)$,
and $p\geq 1$ is some fixed natural parameter.
\end{definition}

We can take this Sturm distance, say with $p=2$ for definiteness, as just one of many possibilities.  Further GH type metrics 
on collections of measured metric spaces as well as discussions of their computational implementation in practice for digital shape comparison can be found in the very accessible \cite{gh-shape}.

\section{Configuration spaces}\label{app:b}

As in the text, we have:

\begin{definition}If $X$ is a metric space then a {\it (labelled) configuration} in $X$ of $n\geq 1$ points is a collection of $n$ distinct labelled points in $X$.  Let
$$\Gamma_X^{(n)}=\{ (p_1,p_2,\ldots ,p_n)\in X^n: p_i\neq p_j~{\rm if}~i\neq j\},$$
where the exponent $X^n$ denotes the $n$-fold Cartesian product.
Let $\Gamma_X=\sqcup _{n\geq 1} \Gamma_X^{(n)}$ denote the disjoint union.  
\end{definition}

Insofar as $\Gamma_X^{(n)}$ is a subset of the Cartesian product $X^n$, it inherits a natural subspace topology.
For example, $\Gamma_X^{(1)}=X$ while a configuration on two points is the complement of the diagonal
in the Cartesian square $\Gamma_X^{(2)}=X^2-\{ (x_1,x_2)\in X^2:x_1=x_2\}$. 
 
There is a nice theory of analysis and measure on configuration spaces 
satisfying appropriate boundedness conditions, cf.\ \cite{AKR98a}.  For our purposes here,
just the underlying topological spaces and vector bundles over arbitrary configurations are enough.

\begin{definition} Given not only a metric space $X$ but also a Hilbert space $V$, we define
$$E_X^{(n)}=\{ \bigl ((p_1,\varphi_1),(p_2,\varphi_2),\ldots ,(p_n,\varphi_n)\bigr)\in (X\times V)^n:
p_i\neq p_j~{\rm if}~i\neq j\},$$  and set $E_X=\sqcup_{n\geq 1} E_X^{(n)}$.
\end{definition}

Again, as a subset of the Cartesian product $(X\times V)^n$,
the new space $E_X^{(n)}$ inherits its natural metric topology, and the forgetful map 
$E_X^{(n)}\to\Gamma_X^{(n)}$ has the natural structure of a 
vector bundle with fiber $V^n$ for each $n\geq 1$.  We informally
refer to $E_X$ itself as a $V$-bundle over $\Gamma_X$.

Taking $X={\mathbb R}^3$ and $V=[L^2({\mathbb S}^2)]^{N+1}$ gives a vector bundle
$E=E_X$ over $\Gamma=\Gamma_X$, whose fiber is the state of the organism.


\begin{thebibliography}{ABCD}


\bibitem{AKR98a} S.\ Albeverio, Yu.\ G.\ Kondratiev,  M.\ R\"ockner. Analysis and geometry
on configuration spaces. J. Funct. Anal. {\bf 154} (1998), 444-500; Analysis and geometry
on configuration spaces: The Gibbsian case. J. Funct. Anal. {\bf 157} (1998), 242-291.

\bibitem{Watson}  B.\ Alberts, A.\ Johnson, J.\ Lewis, M.\ Raff, K.\ Roberts, and P.\ Walter,
Molecular Biology of the Cell 4th Edition, Garland (2002).


\bibitem{birman} J.\ Birman, Braids, Links and Mapping Class Groups, 
Annal of Mathematical Studies {\bf }, Princeton University Press (1974).

\bibitem{izzy} Y.\ Katznelson, 
An Introduction to Harmonic Analysis,
Cambridge University Press (2004).

\bibitem{lobo} D.\ Lobo, W.\ S.\ Beane, M.\ Levin,
Modeling Planarian Regeneration: A Primer for Reverse-
Engineering the Worm, PLoS Computational Biology {\bf 8}
(2012).

\bibitem{Levin} M.\ Levin,
Morphogenetic fields in embryogenesis, regeneration, and cancer: Non-local
control of complex patterning, BioSystems {\bf 109} (2012), 243-261.

\bibitem{gh-shape} F.\ M\'emboli, On the use of Gromov-Hausdorff distances for
shape comparison, Europgraphics Symposium on Point-Based Graphics (2007),
M.\ Botsch, R.\ Pajarola (editors).


\bibitem{MS} N. Morozova and M. Shubin,
The Geometry of Morphogenesis and the Morphogenetic Field Concept 
(2012).

\bibitem{Stu06} K.-T.\ Sturm, On the geometry of metric measure
spaces. I. Acta Math. {\bf 196} (2006), 65-131.

\bibitem{thom} R.\ Thom, Structural stability and morphogenesis,
W.\ A.\ Benjamin (1972).




\end{thebibliography}
\end{document}